\documentclass[aps,prx,twocolumn,groupedaddress,showkeys,10pt]{revtex4-2}

\usepackage{braket}
\usepackage{graphicx}
\usepackage{placeins}
\usepackage{amsmath}

\begin{document}

\title{First Principles Magnetic Initialisation Made Simple}

\author{Kit D. Brown}
\affiliation{School of Physics, Engineering, and Technology, University of York, York, YO10 5DD, UK}

\author{Robert A. Lawrence}
\affiliation{Department of Physics, University of Tampere, S\"{a}hk\"{o}talo, Korkeakoulunkatu 3, 33720 Tampere, Finland}
\email{robert.lawrence@tuni.fi}

\date{\today}

\begin{abstract}

A fundamental issue with first principles electronic structure calculation is the requirement for magnetic initialisations, which are usually derived from experimental results. In this paper, we present a new and computationally inexpensive method to predict magnetic structures from first principles, based upon symmetry analysis and a non-magnetic band structure only. We demonstrate that this method of creating initialisations successfully reproduces the known magnetic ground state for a wide class of magnetic materials, including ferromagnetic BCC Fe, antiferromagnetic L1$_0$ PtMn, and frustrated antiferromagnetic L1$_2$ IrMn$_3$, while also predicting non-magnetic materials to remain non-magnetic.
\end{abstract}


\maketitle

\section{Introduction}
Magnetism has been of interest within density functional theory for nearly as long as the field has existed, with spin being explicitly considered as a fundamental degree of freedom for paramagnetic materials even in the original Kohn-Sham paper \cite{Kohn1965} on which modern practical calculations are built. In principle, permanent magnetism is an emergent property of the electronic system and the correct magnetism-aware ground state (whether magnetised or not) should ``just emerge'' from optimising towards the electronic ground state. Unfortunately, in practical simulations spin does not ``just emerge'' due to the use of local optimisation algorithms within the self-consistent field (SCF) procedure that fail to break the global symmetry of the non-magnetic state. Nevertheless, magnetic materials have been of ongoing interest within electronic structure theory and a workaround technique has been developed; magnetic initialisation based on treating spin, incorrectly, as an input variable to the simulation. 

Magnetic initialisation, also known as spin initialisation, works by inputting a large \cite{Wong2026} initial magnetic moment corresponding to the correct magnetic ground state texture. This forces the system into the local minimum associated with the desired magnetic state, causing the system to converge correctly. This can either be done through a global optimisation for magnetic structure, causing the expense of the calculation to rise significantly and risking missing antiferroic order that may not be compatible with the initial model of the system, or else through adding information from experiment, sacrificing the first-principles robustness of electronic structure methods in favour of a fast, empirical method that gives the desired answer. 

In this paper, we present a method for magnetic initialisation that can determine a minimal set of magnetic ground states from first principles in a computationally efficient, systematic and robust manner, including detecting magnetic order larger than an initially provided primitive cell. 

\section{Background Theory}

Our method draws together representation theory, the first and second Jahn-Teller and hidden Jahn-Teller effects, including the extensions of those to band theory, and the epikernel principle of Ceulemans \emph{et al.} \cite{Ceulemans1984}. 

Firstly, we will recap Jahn-Teller theory and demonstrate its validity for magnetisations in addition to the well-known case of atomic displacements, before proceeding to set out our method. Further details on representation theory and the epikernel principle, including the mathematical procedures for executing our method are found within appendices \ref{app:A} and \ref{app:B}.

\subsection{Jahn-Teller Theory and Magnetism}
The original Jahn-Teller \cite{JT1937} proof proceeded through elegant group-theoretical arguments. Starting with the energy correction from first order perturbation theory
\begin{equation}
    \Delta E = \bra{\psi_0}\hat{H}'\ket{\psi_0}
\end{equation}
they observed that the only way for an integral over all space to be non-zero is if the representation of the integrand contains the totally symmetric irreducible representation (irreps). Beginning with point group irreps, appropriate to their limited case of molecules, they determined the representation of the integrand by considering the direct product of the representation of its components. This is a commutative property, and so one may freely determine the direct product of the wavefunction components first. 

For states corresponding to single-dimensional irreps (A and B irreps in Mulliken notation), their direct product consists of, and only of, the totally symmetric irrep. Accordingly, the only way for $\Delta E$ to be non-zero is if the perturbation $\hat{H}'$ is also totally symmetric. By definition, a totally symmetric perturbation cannot break symmetry, making the case of single-dimensional irreps somewhat less exciting. 

For multi-dimensional irreps, that direct product does contain the totally symmetric irrep, but also other irreps as well. Now, in order to get a totally symmetric component, symmetry-breaking perturbations that transform as one of those other irreps are also valid. In this case, the degeneracy will be lifted, with its irrep subducing appropriately in the final point group of the crystal. Lifting the degeneracy requires the introduction of a splitting, and if the electron orbitals are only partially occupied this causes a total reduction in the energy of the system.

Furthermore, we note that work considering atomic movement usually only considers the totally symmetric component of the direct product. This is because atomic motion (linear momentum) is a vector quantity, and therefore intrinsically transforms symmetrically. Angular momenta, such as spin, transform as pseudovectors, which are inherently antisymmetric physical objects. By allowing the antisymmetric component to interact with pseudovector perturbations, such as magnetisation, we find that magnetisation is intrinsically covered by Jahn-Teller physics. 

Before moving to extensions to the Jahn-Teller theory, we note that only perturbations with irreps which correspond to normal modes of the crystal (e.g. phonons) may be \emph{spontaneously} symmetry breaking. For other irreps, symmetry-breaking may only be achieved by application of an external field, as is the case for strain-engineering.

\subsection{Hidden and Second-order Jahn Teller Effects}

We also note that this argument need not apply only to the first order contribution to the total energy; it also applies at higher orders (referred to as the ``pseudo Jahn-Teller'' interaction) \cite{Bersuker2020}.
At second order,
\begin{equation}
    \Delta E = \bra{\psi_0}\hat{H}''\ket{\psi_0} + \frac{\bra{\psi_0}\hat{H}'\ket{\psi_1}}{E_{1}-E_{0}}
\end{equation}
and the same symmetry arguments about the integrand of \emph{either or both} parts containing the totally symmetric irrep still apply. This is commonly understood to be the mechanism underpinning conventional ferroelectric materials \cite{Spaldin2026}.

To complete our brief recap of the Jahn-Teller family of effects are the \emph{hidden} effects \cite{Bersuker2023}. Unlike the conventional Jahn-Teller effect, the degeneracy occurs in an unoccupied state. If the state is sufficiently close to the Fermi level, and the perturbation is sufficiently large, the lower-energy branch of the lifted degeneracy can become lower energy than the high-symmetry valence states leading to a total reduction in energy. The practical effect of which is that a \emph{large} perturbation rather than an infinitesimal one is required to stabilise the system. We note in passing that this explains the conventional wisdom that magnetic initialisations should use large moments not small ones in order to lead to successful convergence to a magnetic state. We direct the interested reader to the work of Isaac Bersuker \cite{Bersuker2020,Bersuker2020b,Bersuker2023}.

\subsection{Band Jahn-Teller Theory}

For crystalline cases, we should also consider the band Jahn-Teller effect \cite{Polinger2007}, strictly a type of pseudo-Jahn-Teller effect. Therefore, taking only the term that is first order in the Hamiltonian, we have
\begin{equation}
    \Delta E = \frac{\bra{\psi_{0,\vec{k}}}\hat{H}_{\delta \vec{k}}'\ket{\psi_{0,\vec{k}'}}}{E_{0,\vec{k}'}-E_{0,\vec{k}}}
\end{equation}
Now, the symmetry arguments proceed as before, but with the additional constraint of conservation of crystal momentum; $\delta \vec{k} = \vec{k}'-\vec{k}+\vec{G}$, where $\vec{G}$ is any reciprocal lattice vector. We note that this generalisation also requires the use of space groups rather than simply crystallographic point groups. 

For $\delta\vec{k}\neq 0$, we observe that the addition of a single mode with finite crystal momentum implies a time dependent ground state. In order to construct a time-independent ground state, we observe that the selection rule $\delta \vec{k} = \vec{k}-\vec{k}'+\vec{G}$ is also simultaneously valid. Adding perturbations corresponding to these wavevectors simultaneously yields a standing wave solution, giving a time independent ground state.

\subsection{Epikernel Principle and Limited Symmetry Descent}

Whilst the Jahn-Teller theorem lets us identify unstable states from a band structure alone, it does not provide a natural stopping point; yet we know that the vast majority of physical systems in their ground state correspond to highly-symmetric cases. A rigorous formalism for this is provided by the epikernel principle of Ceulemans \emph{et al.}, which states that given an unstable high symmetry parent group, the extremal points in the PES correspond to maximal epikernel subgroups of the original group. By identifying these extremal points (which may be maxima or minima), we can identify a minimal set of perturbations that are both symmetry-breaking in the initial group and totally symmetric within the maximal epikernels. This may be achieved through use of the standard projection formula (see Appendix \ref{app:B}).  

\begin{figure}
    \centering
     \includegraphics[width=\linewidth]{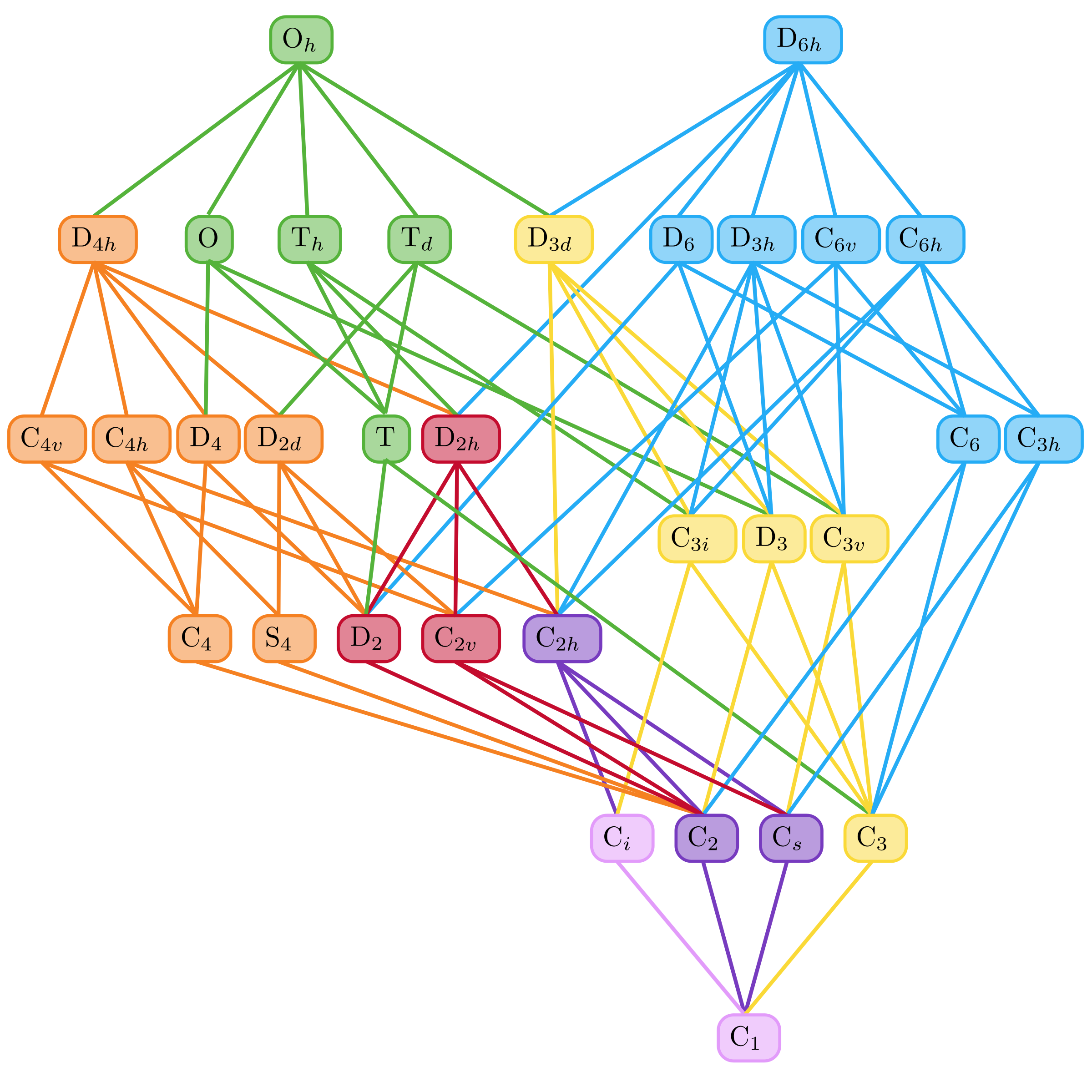}
    \caption{Descent of symmetry graph depicting the group-subgroup relations for the crystallographic point groups using Sch\"onflies notation. This maps drectly to Translationengleiche subgroups of space groups. All groups are subgroups of at least one of the D$_{6h}$ or O$_h$ point groups, and are supergroups of the trivial group, C$_1$ that consists only of the identity element. The colour of each group label indicates the kind of crystal system that has the symmetry: cubic systems are shown in green, hexagonal in blue, tetragonal in orange, trigonal in yellow, orthorhombic in red, monoclinic in purple, and triclinic in lilac. Subgroups are always shown below their supergroup.}
     \label{fig:descent}
\end{figure}

In a more qualitative sense, one may explain the epikernel principle thusly; having identified valid irreducible representations, $\Gamma_i$ of the initial group, $\mathcal{G}$ it is possible to perform a symmetry-descent such that the initial representation may only be expressed as the totally symmetric representation of the subgroup. This is given the somewhat grand name of kernel of $\mathcal{G}$ with respect to $\Gamma_i$, and may be thought of as containing all the symmetries that can never be lifted by a perturbation that transforms as $\Gamma_i$. It is also possible to construct the group of symmetries that \emph{may} be broken, known as the homomorphic image \cite{Ceulemans1984}. Minimal symmetry breaking consists of moving from the homomorphic image, $\mathcal{H}$, to one of its direct subgroups, $\mathcal{S}_H$. Applying the inverse of the mapping used to construct $\mathcal{H}$ originally now leads to a \emph{supergroup} of the kernel. This supergroup, in order to obey linguistic best-practice, gains the fully Greek name of an \emph{epikernel}, and represents a partially broken symmetry of $\mathcal{G}$. If no additional epikernels exist in the symmetry descent chain (see figure \ref{fig:descent}) between the original group, $\mathcal{G}$, and a particular epikernel, then that is termed a ``maximal epikernel''. The process to determine a maximal epikernel is given in Appendix \ref{app:A}.

\subsection{Initialisation Workflow}

We now have all the pieces required to perform our method for magnetic initialisation. This method proceeds as follows:

\begin{enumerate}
    \item Evaluate a strictly non-magnetic band structure sampling all high-symmetry points and lines.
    \item Identify potential degeneracies and intraband pseudo-degeneracies within a window either side of the Fermi level.
    \item Evaluate the antisymmetric irrep contained within the direct product of representations of degenerate states
    \item Confirm that irrep corresponds to a valid pseudovector mode of the crystal
    \item Determine maximum epikernels associated with that irrep and the original space group of the crystal
    \item Create symmetry functions of that irrep with appropriate crystal momentum which lead to each separate maximum epikernel
    \item Evaluate the energy of the system following initialisation with each symmetry function
\end{enumerate}

We note that the antisymmetric term only appears in the decomposition of the direct product of the irreps of the states if and only if the two states transform as the same irrep of the same group. This is a significant restriction that vastly reduces the number of possible combinations of degenerate or pseudodegenerate states that may lead to magnetisation. This is a physical restriction on when a high-symmetry system is unstable with respect to magnetisation, and helps to explain the relative paucity of magnetoelectric systems \cite{AHill2002,Hill2000} where both order parameters are ferroic \cite{Spaldin2026}; separate degeneracies are needed to drive each independent order parameter, leading to a direct competition between the mechanisms. We also observe that in the case of \emph{chiral} systems, the antisymmetric component of the direct product of degenerate states may also appear as a symmetric component. In this case, hybridisation between the two (as e.g. magnetophonons) may provide an alternative pathway to intrinsic multiferroism.

It is also worth noting that in addition to the point-group symmetry reduction leading to the Translationengleiche subgroups, which is familiar in the Jahn-Teller molecular literature, the extension to crystals also enables breaking of translational symmetries, leading to the Klassengleiche subgroups. All need to be considered in order to have full coverage of potential symmetry-lowering mechanisms.

\section{Methods}
Initial geometry optimisation, band structure and total energy calculations were performed using CASTEP 26.1 \cite{CASTEP}, using the LDA functional and the QC5 pseudopotential library with cut-off energy of 500 eV and a 5$\times$5$\times$5 Monkhorst-Pack sampling of reciprocal space \cite{MPgrid}. Band structures were sampled non-self consistently with a sampling density along the path of 0.01\AA$^{-1}$. These are deliberately converged to a very coarse level to highlight that relatively computationally intensive high-quality initial calculations are not required to compute potential initialisations. Similarly, the LDA was chosen as the most computationally-lightweight functional, although we note the method applies equally well for any choice of functional. Since degeneracy has its origins in symmetry, the presence of degeneracies is not affected by choice of functional. Whilst different functionals yield slightly different band structures, the degeneracies present result from the symmetry of the system which remains unchanged across functionals. A different choice of functional will change the \textit{energy} of a degeneracy with respect to the Fermi level, however any semi-local functional has no directional dependence, so will not lead to symmetry breaking. Accordingly, whilst different functionals may predict different global ground state orders, the method we present correctly determines the ground state \emph{for that choice of functional}.

Identification of potential degeneracy was performed by eye, with all degeneracies at or within a window above 5 eV of the Fermi level being considered. We note the size of this window has not been optimised, however for degeneracies that are not linked to bands that cross the Fermi level, this is likely a significant overestimation for a potential stabilisation. For systems where the degeneracy is a maximum for bands that cross the Fermi level, the practical window is likely to be wider, leading to a $5\,$eV window used throughout. 

Evaluation of magnetic structures were performed using a full vector treatment of magnetism, which required use of the SOC19 pseudopotential library, and a standard plane wave cut-off of $1200\,$eV and the same 5$\times$5$\times$5 Monkhorst-Pack grid were used throughout. In all calculations, spin-orbit coupling was neglected. Non-magnetic energies, used for comparison with these, was re-evaluated with the same parameters but with an uninitialised state. Both net spin and net absolute spin were checked in all cases to confirm the final magnetic state.

Symmetry analysis was performed using the spglib \cite{Togo2024} library to determine the initial space group of the system, and the spgrep \cite{Shinohara2023} library used to obtain the irreducible representations belonging to that group. Custom code contained within the supplementary information generates either a vector or pseudovector reducible representation which is used to carry out the double projection procedure outlined in Appendix \ref{app:B}. The point groups of little groups and their allowed degeneracies were additionally determined through use of the REPRES program on the Bilbao Crystallographic server \cite{Aroyo2006, delaFlor2024}.

\section{Results}

To validate our method, the magnetic structure of a range of systems with well-understood magnetic order were evaluated using the method proposed above. These systems are BCC Fe (experimentally FM), PtMn (experimentally a collinear AFM), IrMn$_3$ (experimentally a frustrated AFM), $\alpha$-MnTe (an altermagnet), and bulk Si (non-magnetic). These were chosen to cover a wide range of systems, and the correct magnetic ground state was determined uniquely for all systems. In the case of Fe, the global ground state was predicted to be degenerate with the magnetic hard axis (a very low-lying excited state) due to our energy evaluation neglecting the spin-orbit coupling (SOC) that causes the splitting between these states; an evaluation including SOC will correctly discriminate which of the two is the true ground state. This demonstrates the validity of the method for the first-principles prediction of magnetic initialisation without experimental information or a full search of the magnetic parameter space.

\subsection{BCC Fe}
\begin{figure}[h]
    \centering
    \includegraphics[width=\linewidth]{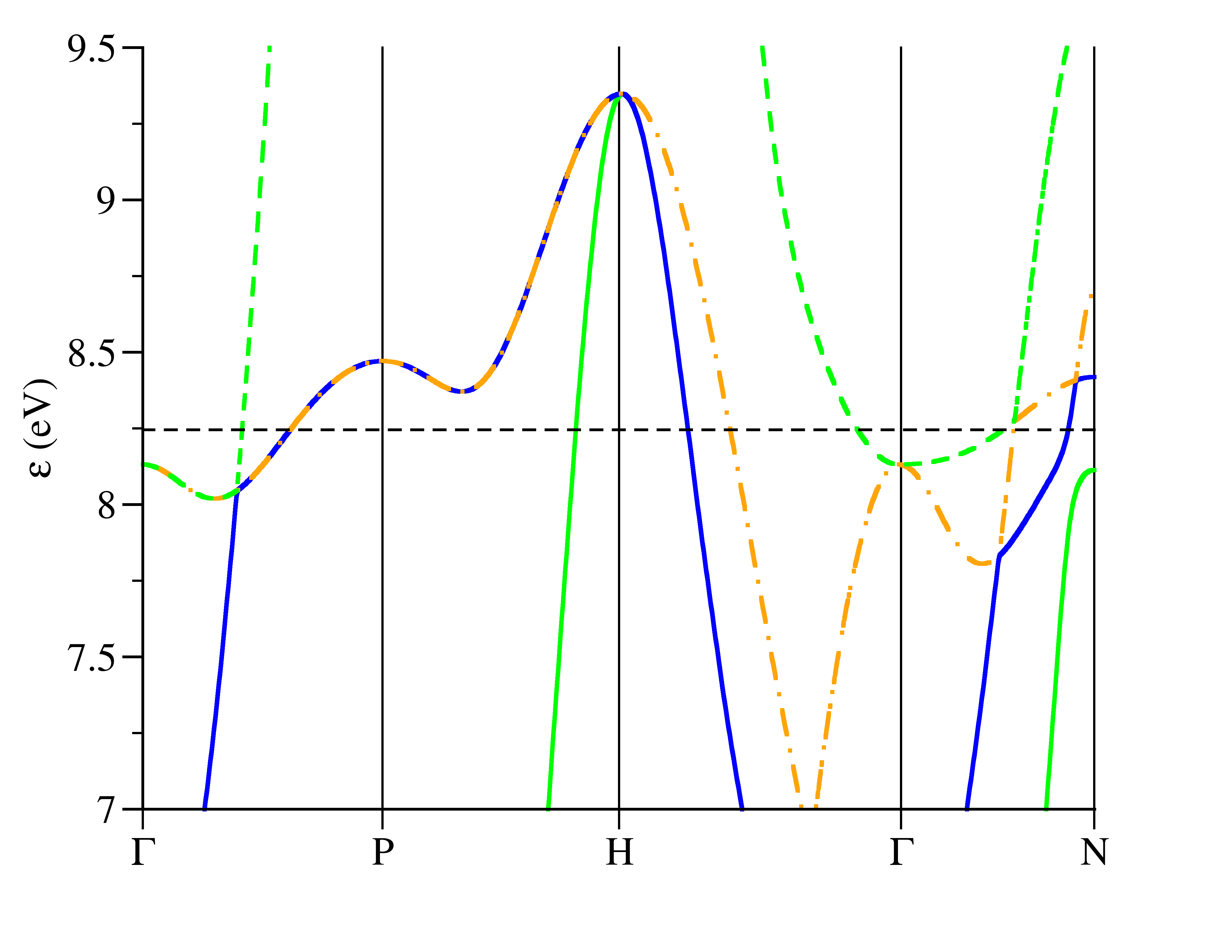}
    \caption{Band structure of non-magnetic BCC Fe. Of particular importance are the doubly degenerate bands between $\Gamma$ and H, indicated by the dashed line, which becomes triply degenerate at H, and is a potential candidate for a hidden Jahn-Teller symmetry-breaking. }
    \label{fig:Fe_NM_BS}
\end{figure}

\FloatBarrier
We begin our examples by considering bulk BCC Fe, which has space group 225, and point group O$_h$. Inspection of the band structure of the non-magnetic primitive cell (figure \ref{fig:Fe_NM_BS}) shows a double-degeneracy between $\Gamma$ and H, which becomes triply degenerate at H. Non-accidental doubly degenerate states belong to a two-dimensional E type irrep, and triply degenerate states must belong to a three-dimensional T type irrep. The antisymmetric component of the direct product of E$\otimes$E is A$_{2g}$ and for T$\otimes$T is T$_{1g}$. Further determination of \emph{which} E or T irrep is chosen is of lesser importance since the direct product with itself is the same regardless of the flavour of E or T irrep chosen. 
The pseudovector modes of the system are expressed wholly in the basis of T$_{1g}$ and T$_{2g}$ and therefore the triple degeneracy is the only potential driving force for spontaneous symmetry breaking. All of the degenerate bands occur at the same k-point and therefore a $\Gamma$-point mode is required to lift the symmetry. The primitive cell contains only one atom (one crystallographic orbit), which combined with the wavevector restriction yields a ferromagnetic configuration.
We also know that the maximum allowed epikernels of T$_{1g}$ pseudovector modes are C$_4$ and C$_3$. The double projection process (see Appendix \ref{app:B}) then yields two sets of options; a ferromagnetic configuration aligned with any one fourfold axis, or a ferromagnetic configuration aligned with any of the threefold axes.

\begin{figure}
    \centering
    \includegraphics[width=\linewidth]{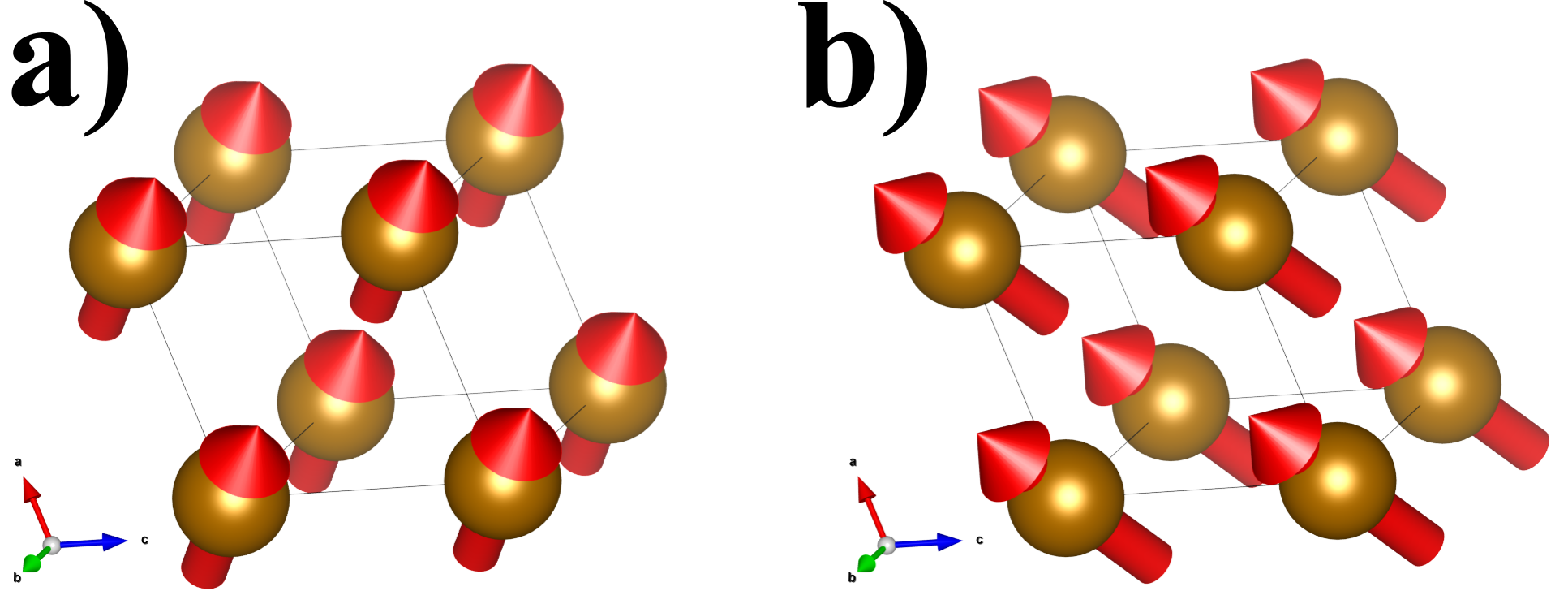}
    \caption{The two magnetic initialisation for BCC Fe. Panel a) is alignment with the 3-fold rotation axis (C$_3$), and b) with the fourfold rotation axis (C$_4$). In the absence of spin-orbit coupling, both are stable and converge to magnetic ground states with the same energy and spins oriented in the same way as their respective initialisations.}
    \label{fig:Fe_spins}
\end{figure}

Since both options correspond to maximum epikernels, both FM (C$_3$) and FM (C$_4$) options (see figure \ref{fig:Fe_spins}) were tested, and the energy of these relative to the non-magnetic system are reported in table \ref{tab:Fe}. These do not show a difference between the two magnetic states; this is in agreement with experiment at the level of theory used to evaluate the energies of the states, which only pick up a small $\sim 1\,\mu$eV\cite{Trygg1995} splitting due to spin-orbit coupling effects that were neglected in these simulations. 

\begin{table}
\begin{tabular}{c|c}
System &  E/eV \\\hline
NM  & 0 \\
FM (C$_3$) & -0.22 \\
FM (C$_4$) & -0.22 \\
\end{tabular}
\caption{Energy difference between non-magnetic (NM) and the predicted FM configurations. The energy difference between easy (FM C$_4$) and hard (FM C$_3$) magnetic orientations are 0 to within the convergence of the calculations, as expected given the $\mu$eV scale of the magnetocrystalline anisotropy of Fe. Configurations are shown in figure \ref{fig:Fe_spins}.\label{tab:Fe}}

\end{table}
\FloatBarrier
\subsection{PtMn}
\FloatBarrier

\begin{figure}
    \centering
    \includegraphics[width=\linewidth]{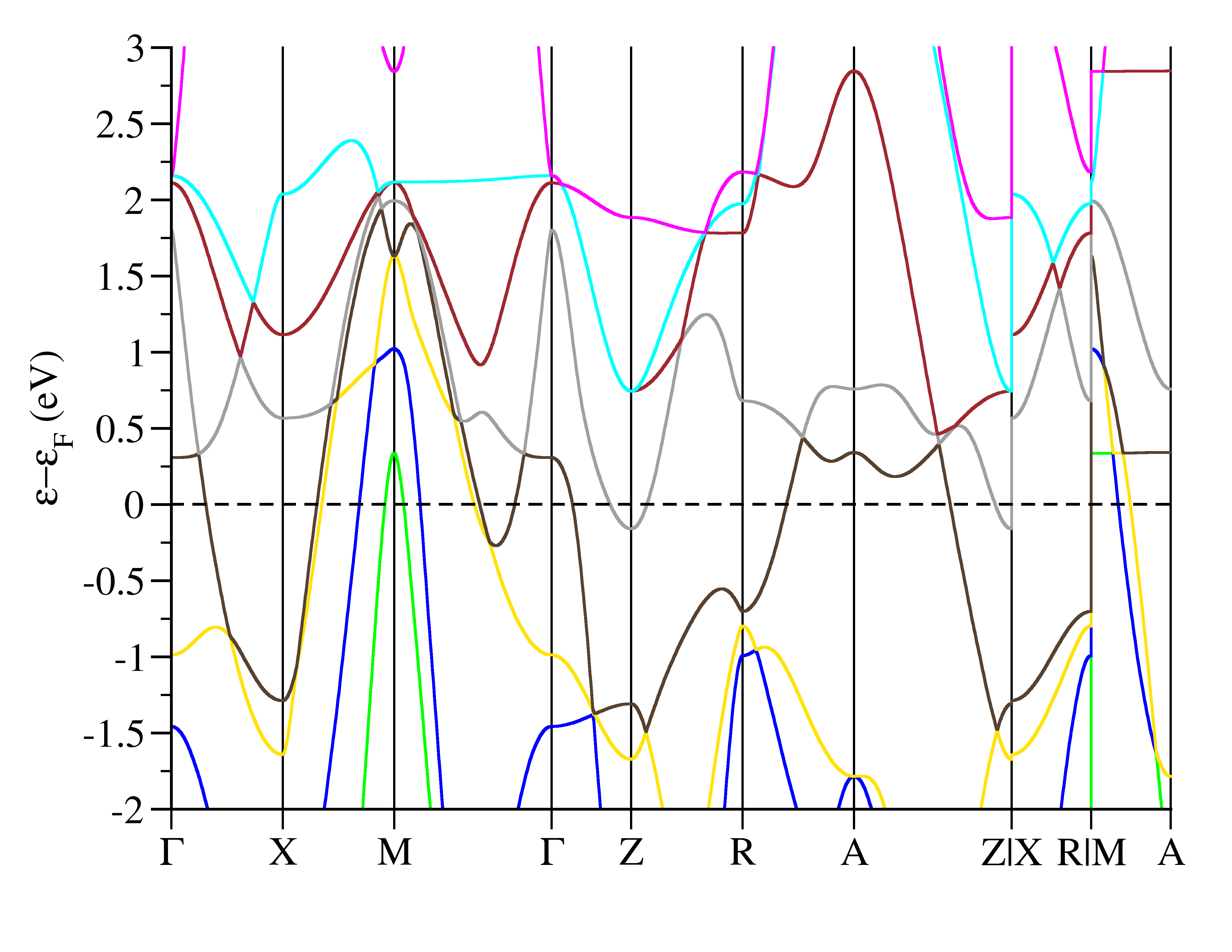}
    \caption{Band structure of the primitive cell of non-magnetic L1$_0$ PtMn. The black horizontal dashed line at $\sim$9 eV is the Fermi level. Of particular importance is the cyan band found at $\sim 2$\,eV above the Fermi level between $\Gamma$ and M, at which points it is doubly degenerate. The cyan band is also degenerate with the brown band along most of the high symmetry line $\Gamma \rightarrow$ Z. }
    \label{fig:PtMn_NM_BS}
\end{figure}

PtMn is, experimentally, a collinear antiferromagnet with an FCT (L1$_0$) crystal structure \cite{Solina2019}, space group 123 (P4/mmm). This magnetic structure is incommensurate with the non-magnetic primitive cell, which contains only 1 formula unit. The band structure for this primitive cell is shown in figure \ref{fig:PtMn_NM_BS}.   Its point group, D$_{4h}$, is capable of supporting doubly degenerate states (it has 2 2-D irreps), but only accidental triply degenerate states. Away from the zone centre and zone edges, only the LD (0,0,X) and V (0.5,0.5,X) high-symmetry lines are capable of supporting degenerate modes for arbitrary values of X.

The key degeneracies for the non-magnetic PtMn primitive cell occur at $\Gamma$, M, Z, and along the $\Gamma \rightarrow$ Z line. Consideration of the exact degeneracies directly requires that the perturbation has a wavevector of 0, transforming as an irrep of the space group at the $\Gamma$ point, however we note that the small energy separation between $\Gamma$ and M of about 40 meV makes this a likely candidate for pseudodegeneracy corresponding to a wavevector of ($\frac{1}{2}$,$\frac{1}{2}$,0). The direct product E $\,\otimes\,$ E = $A_{1g}\oplus [A_{2g}]\oplus B_{1g} \oplus B_{2g}$ has the A$_{2g}$ irrep as its antisymmetric component regardless of whether the modes are both $E_g$ or $E_u$. As the same band connects the pseudodegeneracies as $\Gamma$ and M, the two degeneracies must be both gerade (E$_g$) or both ungerade (E$_u$), leading to the same symmetry selection rules.

At this point, we observe that for a wavefunction where spin and spatial components are separable, we have two potential sources of splitting; spin-splitting and ``spatial-splitting''. For the spin-polarised case, we \emph{do not} require that the non-magnetic bands (spatial-only component) is split by the perturbation since the \emph{spin} component is split. For the non-spin-polarised case (such as antiferromagnets), the representation of the bands \emph{must} subduce such that the representation of the spatial component is reducible within the final subgroup.

The maximum epikernel of D$_{4h}$ under a pseudovector perturbation corresponding to the $A_{2g}$ irrep is C$_4h$. The C$_{4h}$ group is not compatible with spontaneous symmetry breaking through a reduction in spatial symmetry as E$_g$ subduces to E, with the result that this may only lift degeneracy by acting on the spin degrees of freedom (FM state). The maximal allowed epikernel with no lifting of spin-degeneracy is C$_{2h}$. The eigenmode transforming as A$_{2g}$ is a magnetisation oriented along the $\vec{c}$ axis of the cell, with a phase factor acquired along translation along lattice vectors for the antiferromagnetic case, which provides the reduction from C$_{4h}\rightarrow$C$_{2h}$. This reduction may be more readily seen by applying an arbitrary global translation such that the Pt is at the lattice point.

Now turning our attention to the potential pseudodegeneracy between $\Gamma$ and M, we find the 40 meV splitting  is the approximate energy range of a magnon excitation \cite{Kepaptsoglou2025}, and therefore in addition to the Jahn-Teller interactions, an intraband Jahn-Teller interaction along the $\Gamma\rightarrow$M wavevector is a probable candidate. A simple order of magnitude test is used as evaluating the magnon spectrum directly requires knowledge of the magnetic structure, so is not computationally efficient to use as part of determining an appropriate magnetic initialisation of the system. 

Since the psuedodegeneracy is still E$\otimes$E, the same antisymmetric component is valid. Accordingly, C$_2h$ is now the maximum allowed epikernel, and the magnetic initialisation still points along $\vec{c}$. As for the periodicity (conservation of crystal momentum requirement), it is worth noting that for an arbitrary k-point $\vec{k}=(u,v,w)$ the character of identity changes from 1 (or the identity matrix for multi-dimensional irreps) to \begin{equation}
    e^{2\pi i(t_a\cdot u+t_b\cdot v+t_c\cdot w)} 
\end{equation} 
where $t_a,\;t_b$, and $t_c$ are translations along real space lattice vectors. Translation by an integer number of lattice vectors therefore adds a phase-like term to the magnitude of the moment. In the case of the M point ($\frac{1}{2}$,$\frac{1}{2}$,0), this corresponds to no change under a translation of one lattice vector along z, but a sign flip on translation upon any odd combination of translations along the in-plane lattice vectors. Away from integer translations, there is no atom and therefore the magnitude effect can be neglected since spin is initialised on atomic sites only in most DFT packages. We also note that a stationary ground state requires the addition of both the $\Gamma\rightarrow$M and $\Gamma\rightarrow$-M wavevectors. This also removes the complex component of the magnitude when considering wavevectors that do not involve a perturbation with $\vec{q}$ at zone centre or zone edge. Back transforming from momentum space it can be readily seen that a 2$\times$2$\times$1 unit cell is required to represent a structure with this wavevector (AFM$_{xy}$ configuration).

Finally at \emph{all} wavelengths, there is a degeneracy between a point some fraction, $\frac{1}{\lambda}$ along the  $\Gamma\rightarrow$Z line and the $\Gamma\rightarrow -$Z line. This would correspond to an arbitrary wavevector perturbation which transforms as the A$_2$ irrep of the little group (point group C$_{4v}$). This state could exist in a 1$\times$1$\times\lambda$ supercell, and we elect to test this with a $\lambda=2$ supercell. Once more, the C$_4$ subgroup is the maximum epikernel, however this time the sign of the spin flips along a translation along $\vec{c}$ (AFM$_z$ configuration).

\begin{figure}
    \centering
    \includegraphics[width=\linewidth]{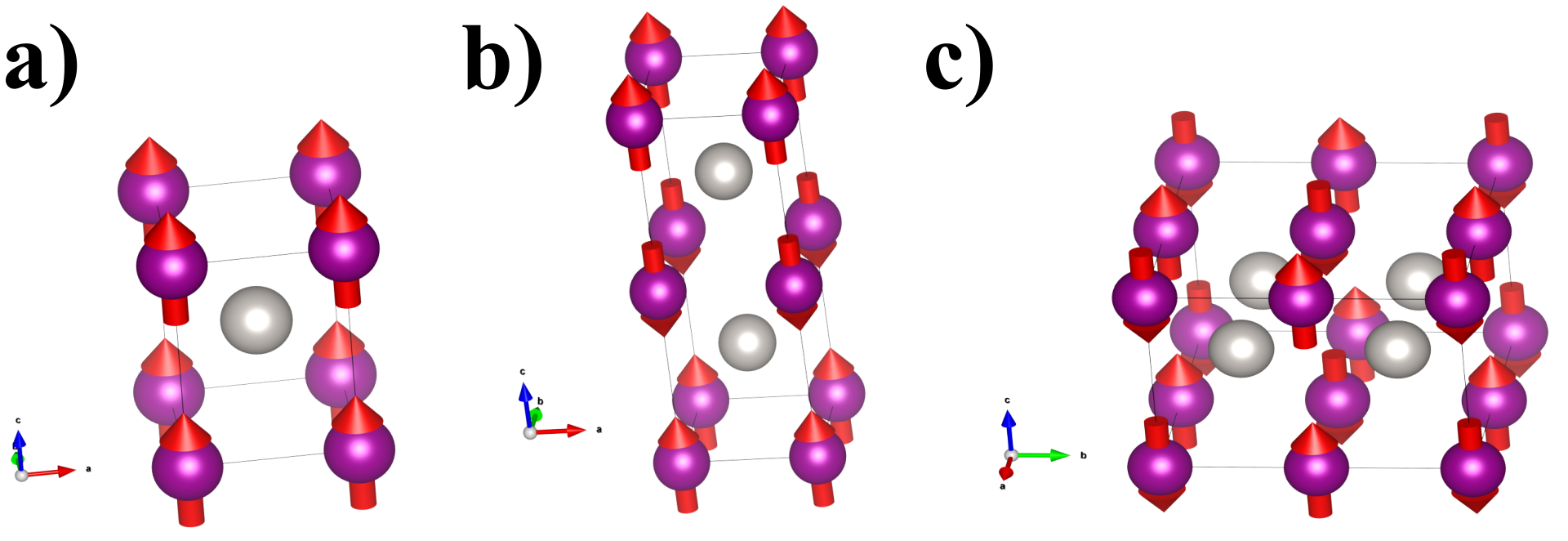}
    \caption{The 3 magnetic initialisations evaluated for PtMn. Panel a) is FM, Panel b) AFM$_z$ and c) AFM$_{xy}$. Note that panel c is a 2$\times$2$\times$1 supercell of the primitive cell and is larger than the conventional cell, which is contained with the in-plane lattice vectors at 45 degrees to those of the primitive cell. Pt atoms are silver, and Mn atoms are purple.}
    \label{fig:PtMn_spins}
\end{figure}

\begin{table}[]
    \centering
    \begin{tabular}{c|c}
      System   & E/ eV \\
      \hline
         NM    & 0  \\
         FM    & -0.05\\
         AFM$_z$ & -0.15 \\
         AFM$_{xy}$    & -0.59
    \end{tabular}
    \caption{Energy difference (normalised to one primitive cell) between non-magnetic (NM) and the predicted FM and AFM configurations of PtMn and the FM configuration. Configurations are shown in figure \ref{fig:PtMn_spins}.}
    \label{tab:PtMn}
\end{table}

These states are shown in figure \ref{fig:PtMn_spins}, and their relative energies normalised \emph{per formula unit} are shown in table \ref{tab:PtMn}. We find that all magnetic phases are more stable than the non-magnetic phase, and the known experimental ground state is the lowest of these three candidate initialisations. Additionally, we note that the correct magnetic initialisation can be determined purely from the primitive unit cell \emph{despite the primitive cell being incommensurate with the correct initialisation}. This is a demonstration that our method offers significant computational efficiency over untargeted global magnetic structure searches, since it can discover incommensurate initialisations without requiring many different unit cell sizes to be searched. We note this aspect can also be applied to non-magnetic perturbations, potentially providing a robust route to ensure that ground state crystal structure is captured, even for charge-density wave ground states.

\FloatBarrier
\subsection{IrMn$_3$}
\FloatBarrier

\begin{figure}
    \centering
    \includegraphics[width=\linewidth]{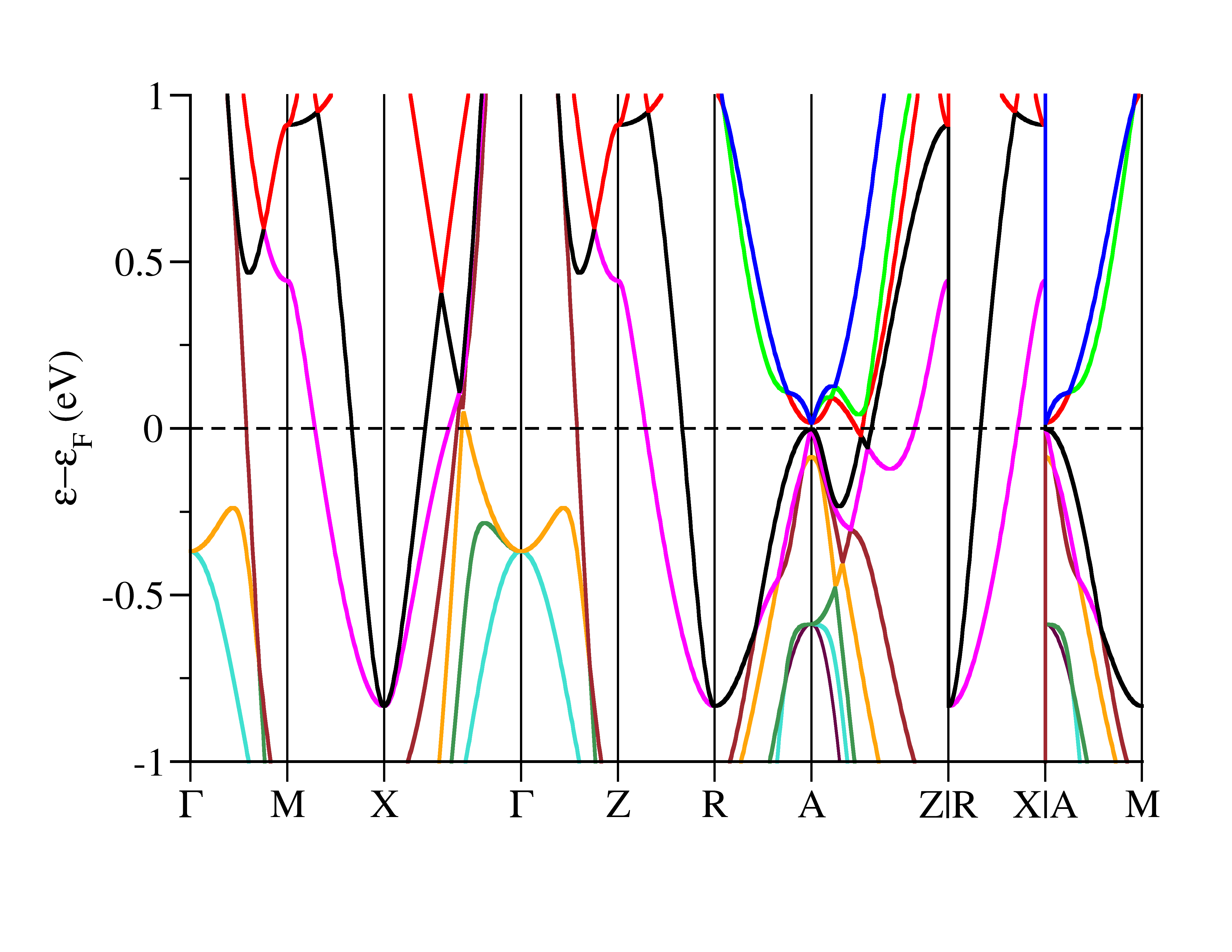}
    \caption{Non-magnetic band structure of IrMn$_3$. This contains a triple degeneracy at the A point, and two double degeneracies along A$\rightarrow$M and A$\rightarrow$R, approximately $\frac{1}{39}$ of the way along those high-symmetry lines. The state associated with lifting these double degeneracies would require a 39$\times$39$\times$39 unit cell and was neglected in this work. Ir atoms are beige and Mn atoms are purple.}
    \label{fig:IrMn_NM_BS}
\end{figure}

The frustrated antiferromagnetic L1$_2$ phase of IrMn$_3$ \cite{Kohn2013} (space group 221 / P$m\bar{3}m$) is widely used in industrial contexts \cite{Szunyogh2009} due to its remarkably high magnetocrystalline anisotropy . Unlike PtMn, its accepted magnetic structure is commensurate with its primitive unit cell. Considering the primitive non-magnetic unit cell's band structure (figure \ref{fig:IrMn_NM_BS}), we can see there is a triple degeneracy either side of the Fermi-level at the A-point, and a set of two double degeneracies either side of A along the A$\rightarrow$M, A$\rightarrow$Z and A$\rightarrow$R lines.  The little group of these high symmetry lines (corresponding to a C$_{4v}$ point group) does not permit triple degeneracy, and close inspection does reveal that there is a very fine splitting here to ensure that requirement is obeyed. A gamma-point lifting of one of these double degeneracies would correspond to a collinear ferromagnetic ground state transforming as the A$_2$ irrep of the C$_{4v}$ point group with the magnetic moments aligned purely along one of the fourfold rotation axes to ensure a final C$_4$ epikernel state.

Since these degeneracies occur at the same energy, a set of intra-band pseudo Jahn-Teller initialisations are also possible. The $\vec{k}\rightarrow-\vec{k}$ degeneracies should all be lifted simultaneously such that the supercell associated with full degeneracy lifting is cubic, despite each high-symmetry line individually being parallel with a single reciprocal lattice vector. The wavelength associated with this ($\vec{q}\sim\big(\frac{1}{39},\frac{1}{39},\frac{1}{39}\big)$) would require a 39$\times$39$\times$39 unit cell to model correctly, and is neglected here due to computational expense and the higher initial energy of this intra-band degeneracy making it less likely to penetrate the non-magnetic many-body state to a greater depth than lifting the A-point degeneracy.   

\begin{figure}
    \centering
    \includegraphics[width=\linewidth]{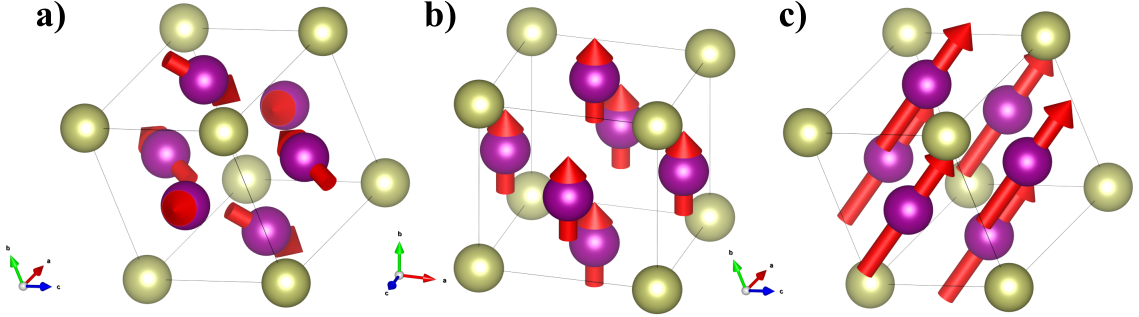}
    \caption{The 3 trial spin configurations of IrMn$_3$. Panel a) is the antiferromagnetic state with a triangular motif, Panel b) is the FM (C$_4$) state with the spins ferromagnetically aligned along one of the fourfold rotations axes, and c) is FM (C$_3$), where the ferromagnetically aligned moments point along one of the threefold rotation axes.}
    \label{fig:IrMn_spins}
\end{figure}

The triple degeneracy at the A-point may be lifted by a T$_{1g}$ mode of the full crystallographic point group with $\vec{q}=0$. This irrep appears multiply in the calculation of pseudovector eigenmodes of the system. For atoms positioned at
\begin{align*}
    \text{Mn}_1:\quad\vec r&=(\tfrac{1}{2},\tfrac{1}{2},0)\\
    \text{Mn}_2:\quad\vec r&=(\tfrac{1}{2},0,\tfrac{1}{2})\\
    \text{Mn}_3:\quad\vec r&=(0,\tfrac{1}{2},\tfrac{1}{2})
\end{align*}
These modes transform as 
\begin{align*}
    f_1 &= \text{Mn}_1:(1,1,0),\quad\text{Mn}_2:(1,0,1),\quad\text{Mn}_3:(0,1,1)\\
    f_2 &= \text{Mn}_1:(0,0,1),\quad\text{Mn}_2:(0,1,0),\quad\text{Mn}_3:(1,0,0)
\end{align*}

for the three equivalent Mn atoms, with all 6 individually being valid T$_{1g}$ modes of the system. We also note that any linear combination of these modes are also valid eigenmodes of the system. The valid maximum epikernels for this excitation are C$_4$ and C$_3$. Unlike previous examples, we now have multiple atoms in the ``Mn'' crystallographic orbit. This additional degree of freedom means that we must consider a third projection, symmetrising not only according to the initial and final space group, but also according to whether the system transforms as the symmetric (ferromagnetic) or antisymmetric (net-moment-free) representations of the time reversal group, $\mathcal{T}$. The symmetric representation contains not only the trivial non-magnetic case, but also all fully-compensated magnetic structures. For the C$_4$ epikernel, we note that only the ferromagnetic case (FM$_4$) is a valid solution, however for C$_3$ both a ferromagnetic (FM$_3$) and antiferromagnetic (AFM) configuration are compatible with the $\vec{q}=0$ magnetisation mode. These states are shown in figure \ref{fig:IrMn_spins}. The total energies of these states are reported in table \ref{tab:IrMn}.

\begin{table}[]
\centering
\begin{tabular}{c|c}
  System   & E/eV  \\
  \hline
    NM  &  0 \\
    AFM &  -0.56 \\
    FM (C$_4$) & -0.04 \\
    FM (C$_3$)  & -0.04\\
\end{tabular}
\caption{Final energies of the different initialisations of IrMn$_3$ with respect to the non-magnetic ground state. In the absence of spin-orbit coupling the energy difference between the two ferromagnetic states cannot be discriminated.}
\label{tab:IrMn}
\end{table}

We note at this stage that many of the local optimisation techniques commonly used in electronic structure optimisation work by purifying eigenstates. If the projection onto the two irreps of $\mathcal{T}$ is not performed, the resulting ``frustrated ferromagnetic" states predicted have still been reduced to the C$_3$ maximum epikernel, but are an admixture of the two irreps of $\mathcal{T}$, FM (C$_3$) and AFM. Standard electron density optimisation techniques can optimise from this state to the correct ground state, in agreement with previous work by one of the authors \cite{Lawrence2026}. We also note that even for IrMn$_3$, a system known for its remarkably high magnetocrystalline anisotropy, spin-orbit effects are relatively unimportant for determining the ground state magnetic structure. We would, however, expect the two FM configurations to be weakly split in energy by spin-orbit coupling effects.
\FloatBarrier
\subsection{$\alpha$-MnTe}

The remaining major class of magnetic materials are altermagnets; which are compensated but show spin splitting. The archetypal altermagnet is perhaps $\alpha$-MnTe, (space group 194). 

\begin{figure}
    \centering
    \includegraphics[width=\columnwidth]{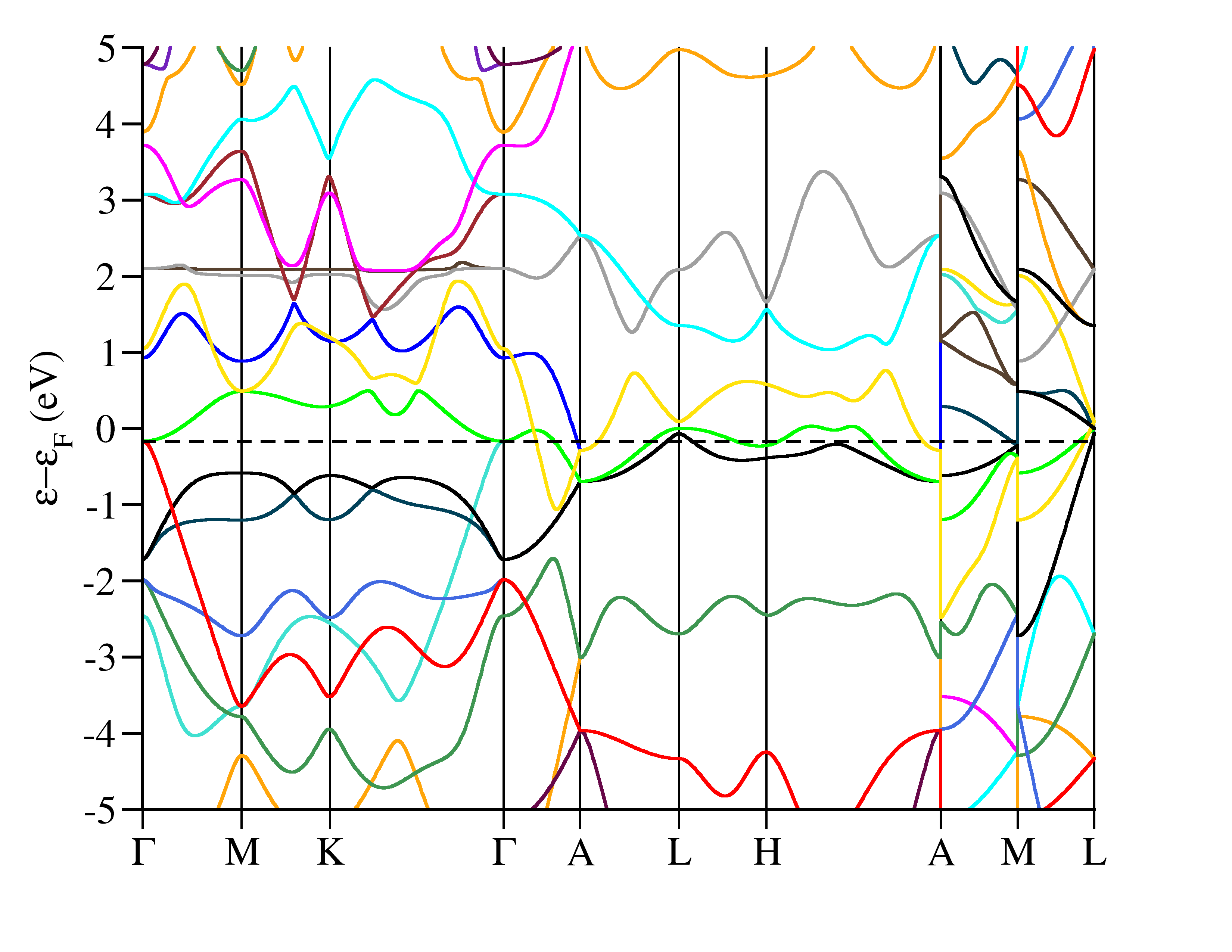}
    \caption{Band structure for $\alpha$-MnTe. The bands with $k_Z=\frac{1}{2}$ are all doubly degenerate, however no pseudo-degeneracies exist between high-symmetry points. Accordingly, an A$_{2g}$ mode with non net momentum ($\Gamma$-point) may act to lift these degeneracies.}
    \label{fig:MnTe_bands}
\end{figure}

Many double degeneracies occur in MnTe, however there are no band-pseudodegeneracies at high k-points. We note that in this case the LDA predicts MnTe to be metallic, however since this does not alter symmetries, just band energies, this underestimate is not a concern for our purposes of spin-structure determination. The $\Gamma$-point pseudovector mode A$_{2g}$ for $\alpha$-MnTe, corresponding to the only possible magnetisation, only has one symmetry function. Whilst the primitive cell contains two separate atoms, these correspond to a single crystallographic orbit. Due to the presence of a screw axis within the space group of $\alpha$-MnTe (space group 194), this causes the two Mn atoms within the basis to have opposite spins, whilst the translation symmetries of the unit cell are preserved, indicating that our spin initialisation method can correctly predict the magnetic order in every class of permanent magnetic material. 

\begin{figure}
    \centering
    \includegraphics[width=0.5\columnwidth]{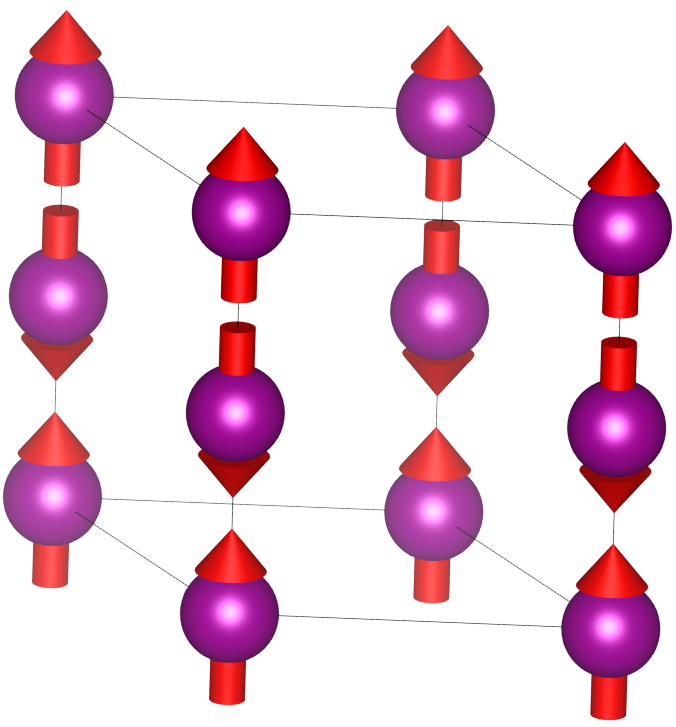}
    \caption{Magnetic order of $\alpha$-MnTe. Whilst transforming as the antisymmetric irrep of the time reversal group, the act of the translation component of the twofold rotation operator about the c-axis (and other elements of the group which include it) act to flip the spin-component, leading to the compensated, but translation-symmetry preserving magnetic order that is characteristic of altermagnets. Since the A$_{2g}$ irrep is single dimensional, only one symmetry function exists.}
    \label{fig:MnTe_strucs}
\end{figure}

\begin{table}[]
\centering
\begin{tabular}{c|c}
  System   & E/eV  \\
  \hline
    NM  &  0 \\
    AM &  -0.78 
\end{tabular}
\caption{Final energy of $\alpha$-MnTe with respect to the non-magnetic ground state. }
\label{tab:MnTe}
\end{table}

\FloatBarrier
\subsection{Si}
\FloatBarrier

\begin{figure}[h!]
    \centering
    \includegraphics[width=\linewidth]{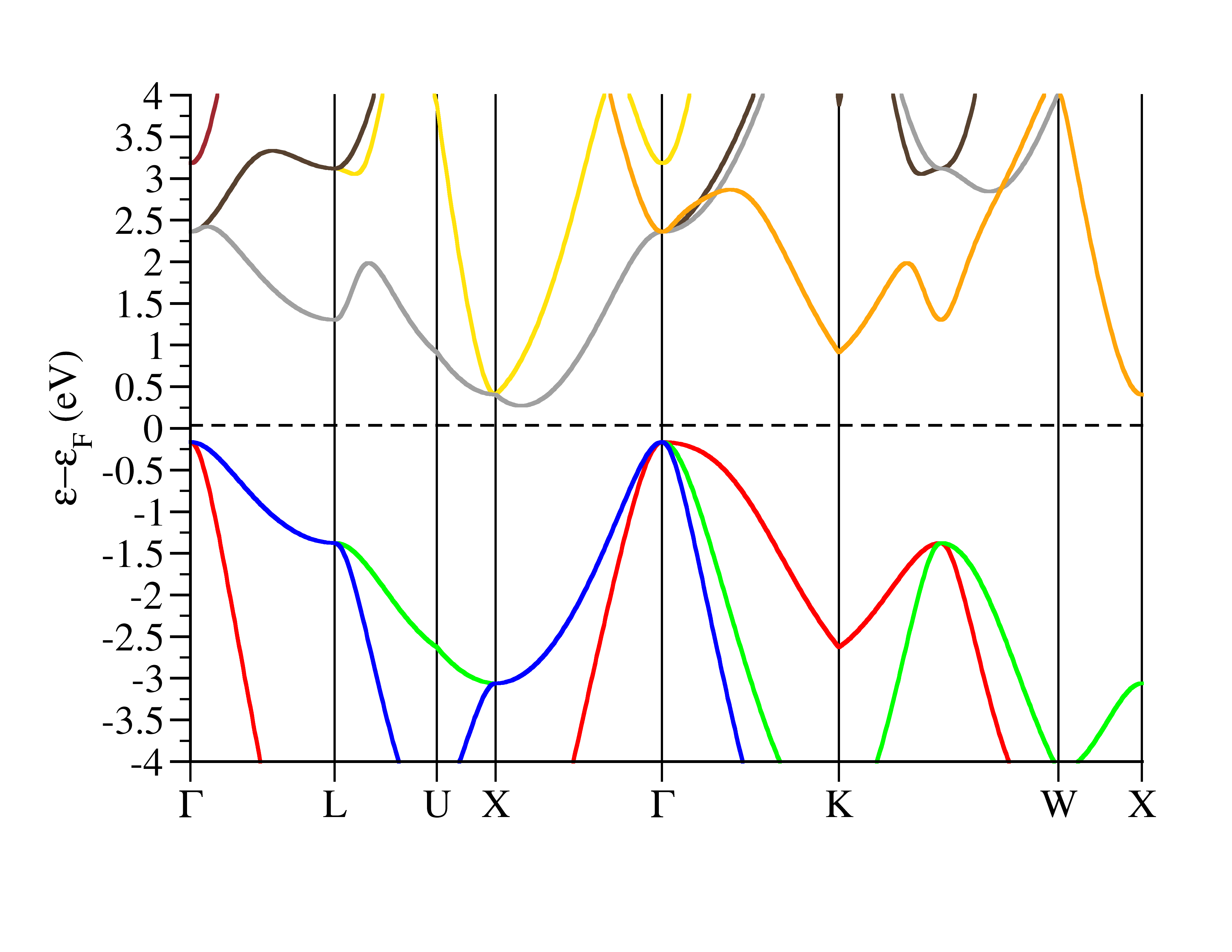}
    \caption{Band structure of non-magnetic Si. A double degeneracy in the excited states exists at X and a much higher energy triple degeneracy exists at $\Gamma$. The horizontal black dotted line is the Fermi level.}
    \label{fig:Si_NM_bands}
\end{figure}

In addition to four well-known magnetic materials, we also considered bulk Si to confirm that are method does not lead to ``false positives''. This system is a non-magnetic insulator in its ground state. A doubly degeneracy exists at the X-point and a triply degenerate state exists at the $\Gamma$-point in the conduction band (see figure \ref{fig:Si_NM_bands}) approximately 2\,eV above the Fermi level. There are no repeats of degeneracies at similar energies at separate k-points indicating only hidden Jahn-Teller instabilities are possible, indicating the only likely magnetic configuration would be an T$_{1g}$ mode with $\vec{q}=0$. The primitive cell contains two silicon atoms, permitting both ferromagnetic and antiferromagnetic configurations.

\begin{figure}[h!]
    \centering
    \includegraphics[width=0.7\linewidth]{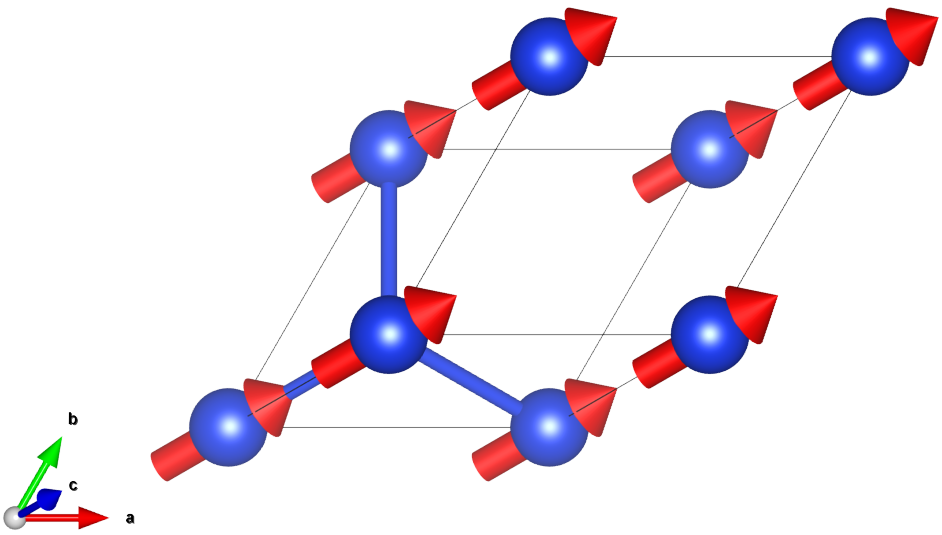}
    \caption{Ferromagnetic initialisation for Si. This is unstable with respect to the non-magnetic case, and self-consistent field optimisation of Si with this initialisation leads to the non-magnetic ground state.}
    \label{fig:Si_spins}
\end{figure}

Unsurprisingly, even when initialised with magnetic moments in this manner (see figure \ref{fig:Si_spins}), the system converges to a completely non-magnetic ground state. The $\sim2$eV direct band gap at $\Gamma$ would require such a large magnetisation to overcome that the partially-occupied p-orbitals alone cannot produce it. This negative result indicates the robustness of this method against falsely predicting magnetic materials as destabilising magnetic initialisations are removed by the self-consistent field process.

\FloatBarrier
\section{Discussion}

We note that the qualitative conclusions of our model are in agreement with the long-standing Stoner model of ferromagnetism \cite{Stoner1938}, however our additional inclusion of band structure information that was computationally inaccessible to Stoner also enables extension to antiferromagnetic orders. Moreover, our model does not require assuming magnetic order \emph{a priori}, a significant advantage to the modern theorist. Unlike the Stoner model, our model's consideration of the hidden Jahn-Teller effect also explains why large spin initialisations are commonly required for the effective single-body system modelled in DFT to produce a magnetic state.

Beyond magnetic materials, we note that in the case of ferroelectric materials, with which the theory of magnetism has long been closely intertwined, the second order (pseudo) Jahn-Teller interaction is widely understood to be the general mechanism \cite{Spaldin2026}. We note that with minor adaptation our initialisation method will also apply to ferroelectrics; although it is probable that consideration of coupling to antiferrodistortive modes will also be critical in many of these cases. Our extension of the Jahn-Teller framework here to apply also to ferromagnetic and antiferromagnetic systems suggests a similar mechanism will apply for antiferroelectrics \cite{Catalan2026}, hinting at a potential single universal framework for understanding all ferroic materials.

\subsection{Note on Axial Phonons}

Axial phonons also carry angular momenta, and therefore may transform as the same irreps as magnetisation modes, which enables them both to hybridise with magnetisations (hence magnetophonons \cite{Nicholas1985}) and co-operatively lower symmetry, or to act on the same degeneracy and competitively lower symmetry. The precise regime depends on the coupling strength (energy separation) between the two modes. It is important to note that in the case of a coupled mode co-operatively lowering symmetry, then our method relies on starting from the high-symmetry case; in this eventuality, starting from an already-lowered crystallographic symmetry presupposes that the displacive (axial phonon) mode competes with the magnetisation to cause symmetry-lowering rather than co-operatively lowering the symmetry.  

\section{Conclusion}

In this paper, we have presented a first principles method capable of predicting magnetic initialisations for density functional theory calculations based on rigorous symmetry considerations. This method uses only the crystal symmetry and a coarse band structure calculation on a non-magnetic system to evaluate the initialisation and is capable of predicting initialisations that are commensurate or incommensurate with the initially-provided unit cell. This offers a more computationally-efficient way to determine magnetic initialisations when \emph{a priori} knowledge is not available from prior studies or experiments. The method was demonstrated to work for BCC Fe, collinear antiferromagnet L1$_0$ PtMn, and frustrated antiferromagnet IrMn$_3$, where the method predicted the correct experimentally known magnetic structure without requiring any reference to experiment. This method is computationally efficient, predicting no more than three candidate magnetic structures per example system, indicating its suitability for widespread use.

\section*{CRediT Statement}
\noindent K.D.B.: Software, formal analysis, writing (review and editing), visualisation, investigation.

\noindent R.A.L.: Conceptualisation, methodology, formal analysis, investigation, writing (original draft), visualisation, supervision.

\appendix
\section{Determining Maximal Epikernels}\label{app:A}

In this appendix, we review the method used for determining maximal epikernels of a group, $\mathcal{G}$, with respect to an arbitrary irreducible representation (irrep), $\Gamma$.

\subsection{Calculate the kernel.}
 Within each group, each irrep spans an orthogonal part of the accessible symmetry space. Accordingly, each irrep leads to a (potentially) different set of minimally-symmetry-broken groups (the maximal epikernels). To find the maximal epikernels, we first determine the kernel of the group under $\Gamma$ by combining all elements of $\mathcal G$ whose character, $\chi$, is identity ($\chi$ equal to the dimension of the irrep) for $\Gamma$ (see table \ref{tab:Ohchar}). This gives the symmetry operations that \textit{cannot} be broken by application of symmetry functions transforming as that irrep. For the $O_h$ point group and the $T_{1g}$ irrep, the kernel, $\text{Ker}(O_h,T_{1g})=\{E,i\}=C_i$. For $E_g$, $\text{Ker}(O_h,E_{g})=D_{2h}$.

 \begin{table}[]
\centering
\begin{tabular}{|c|c|c|c|c|c|c|c|c|c|c|}
\hline
O$_h$ & $E$ & $8C_3$ & $3C_2$ & $6C_4$ & $6C_2'$ & $i$ & $8S_6$ & $3\sigma_h$ & $6S_6$ & $6\sigma_d$ \\\hline
E$_g$ & 2 & -1 & 2 & 0 & 0 & 2 & -1 & 2 & 0 & 0\\\hline
T$_{1g}$ & 3 & 0 & -1 & 1 & -1 & 3 & 0 & -1 & 1 & -1\\\hline
\end{tabular}
 \caption{Rows of the character table for O$_h$ corresponding to the E$_g$ and T$_{1g}$ irreps. The kernel can be determined by selecting the symmetry operations that have character identity under the relevant irreducible representation. For T$_{1g}$ this is $\{E,i\}$ as those are the only two elements with character 3, for E$_g$ this is $\{E,3C_2,i,3\sigma_h\}$ as they all have character 2.}
\label{tab:Ohchar}
\end{table}

\subsection{Calculate the homomorphic image.}
 The next step is to generate the homomorphic image -- the group containing all possible ways of breaking the symmetry -- from the kernel. The homomorphic image is found by calculating all the left cosets of the kernel with the original group. Each of the sets corresponds to a single symmetry operation within the homomorphic image. Using the $E_g$ irrep of $O_h$ as an example here, the first coset is trivially calculated by selecting the identity, $E$ as the symmetry operation from $O_h$:
\begin{multline*}
    \underbrace{E}_{O_h}\cdot\underbrace{\{E,C_2,C_2',C_2'',i,\sigma_h,\sigma_h',\sigma_h''\}}_{\text{kernel}} =\\ \underbrace{\{E,C_2,C_2',C_2'',i,\sigma_h,\sigma_h',\sigma_h''\}}_{\text{coset}}\qquad\rightarrow\qquad E
\end{multline*}

giving $E$ as the corresponding symmetry operation in the homomorphic image. Now selecting another element of $O_h$ that does not appear in the coset, the same process is followed to generate a new coset:
\begin{multline*}
    \underbrace{C_3}_{O_h}\cdot\underbrace{\{E,C_2,C_2',C_2'',i,\sigma_h,\sigma_h',\sigma_h''\}}_{\text{kernel}} =\\ \underbrace{\{C_3,C_3',C_3'',C_3''',S_6,S_6',S_6'',S_6'''\}}_{\text{coset}}\qquad\rightarrow\qquad C_3
\end{multline*}

\subsection{Calculate the Epikernel groups}
    
As we are interested in \textit{minimal} symmetry breaking, we now look to calculate epikernel groups using the map defined by the homomorphic image. These groups describe partially-broken symmetry. To calculate the epikernel groups we consider subgroups of the homomorphic image, and form the epikernel group from the elements of the \textit{cosets} corresponding to the elements within the subgroup. Again considering the E$_g$ irrep of O$_h$, the homomorphic image is D$_3$, which has three subgroups C$_1$, C$_2$, and C$_3$. The epikernel corresponding to C$_1$ is the kernel, D$_{2h}$, C$_2$ maps to D$_{4h}$, and C$_3$ maps to T$_h$. (Applying the mapping to $D_3$ gives the original group, O$_h$). For T$_{1g}$ the epikernel groups are C$_{2h}$, C$_{4h}$, C$_{3i}$, D$_{2h}$, D$_{4h}$, D$_{3d}$, and T$_h$.

\subsection{Determine Maximal Epikernels}

The epikernel groups are the \textit{partial} symmetry breaking groups, while the \textit{maximal} epikernels are the \textit{minimal} symmetry breaking groups. Maximum epikernels can be confirmed by the subduction of $\Gamma$ into the epikernels; the first subgroups along each chain of subgroup descent of $\mathcal{G}$ where the subduction contains the totally symmetric irrep only once are maximal epikernels. The examples of the subduction of the E$_g$ irrep into each of the epikernel groups shown below.
\begin{align*}
    E_g &\overset{D_{2h}}{\longrightarrow} 2A_g\\
    E_g &\overset{D_{4h}}{\longrightarrow} A_{1g}\oplus A_{2g}\\
    E_g &\overset{T_h}{\longrightarrow} E_g\\
    E_g &\overset{D_{4h}}{\longrightarrow} E_g
\end{align*} D$_{4h}$ is the only epikernel group where the fully symmetric representation appears exactly once, and thus is the only maximal epikernel of the $E_g$ irrep. The subduction of T$_{1g}$ into its epikernel groups gives C$_{3i}$, C$_{4h}$, and C$_{2h}$ where the fully symmetric representation appears exactly once, although as C$_{2h}$ is a subgroup of C$_{4h}$ it is \textit{not} a maximal epikernel.

\subsection{Confirmation of Valid Maximal Epikernel}

Confirmation that the degeneracy of the band can be lifted within the maximal epikernel group, the irreducible representation (irrep) corresponding to the symmetry of the band must subduce into multiple components within the target group. We briefly consider the example of a band with E$_g$ symmetry in a system with point group O$_h$ and crystal symmetry T$_{1g}$. As shown in appendix \ref{app:A}, the maximal epikernels are C$_{3i}$ and C$_{4h}$. In C$_{4h}$ the band irrep E$_g$ subduces to $A_g\oplus B_g$, showing the degeneracy is lifted, whereas in C$_{3i}$ E$_g$ subduces to E$_g$, meaning the band stays degenerate and no stabilisation is associated with the corresponding reduction in symmetry hence the C$_{3i}$ state may be neglected.

\section{Determining Symmetry Functions leading to Maximal Epikernels}\label{app:B}

We now review the double-projection method we used to determine the trial perturbations through an initial projection onto the $\Gamma$ irrep of the group $\mathcal G$, followed by a projection onto the fully symmetric irrep of the maximal epikernel group.

\subsection{Calculate the possible symmetry functions for the relevant irreducible representation $\Gamma$ of the original group $\mathcal G$.}

Ordinary symmetry functions, such as may be used to determine magnon or phonon modes through group theory may be found through the use of the projection formula
\begin{equation}
\vec{v}^{\alpha}_m = \sum_{k=1}^{\text{symops}}\chi^\alpha_k(O_k)\bigl[\Gamma(O_k)\vec v\bigr]\;,
\end{equation} where each irrep present in the decomposition of the vector (or pseudovector) reducible representation gives rise to a number of symmetry functions, dependent on both its dimension and multiplicity. For example, a reducible representation with decomposition $A_{2g}\oplus E_g\oplus3T_{1g}$ would have 12 symmetry functions: one from the 1D A$_{2g}$ irrep, two from the 2D E$_g$ irrep, and nine from the three instances of the 3D T$_{1g}$ irrep. (This is the same as the dimension of the reducible representation: for a structure with $n$ atoms, considered in $m$ dimensions there are $n\times m$ orthogonal symmetry functions.) These symmetry functions are found by applying the projection formula onto a set of basis vectors $\vec v$ (chosen to be the unit vectors along each direction of the $n\times m$ dimensional space), with the symmetry operations of the initial group $\mathcal G$, and the $\Gamma$ irrep.

\subsection{Project the symmetry functions onto the fully symmetric space of the maximal epikernel group.}

 However, we note that the significant degrees of freedom this provides can be further restricted by imposing the physical requirement that the perturbation must be fully symmetric in the subgroup of the post-perturbation system (\emph{i.e.} in the maximal epikernel group). If we again consider a crystal with reducible representation $A_{2g}\oplus E_g\oplus3T_{1g}$ in the O$_h$ group, moving to the C$_{3i}$ group, there are only 4 possible symmetry functions that fulfil this requirement, rather than 12. 
 
 This can be seen by considering the subduction of the reducible representation within the C$_{4h}$ group, $A_{2g} \oplus E_g \oplus 3T_{1g} \rightarrow(B_g)\oplus(A_g\oplus B_g)\oplus3(A_g\oplus E_g)$, where the fully symmetric irrep appears once within the decomposition of each multi-dimensional irreps.

  Restricting to the symmetry functions that obeys both constraints is achieved by applying the projection formula a second time, where the group is the maximal epikernel, the irrep is the fully symmetric irrep, and the basis vectors are the symmetry functions found from the first application of the projection formula. This does not fully remove all degrees of freedom of the problem, for example an O$_h \rightarrow $D$_{4h}$ transition in bulk Fe is valid along any of the C$_4$ axes ($\vec{x},\vec{y},\vec{z}$), however the choice will be identical under a global symmetry, such as rotation or translation, and therefore the physical system derived will be identical regardless of choice.

\subsection{Project onto spin-maximising and spin-minimising irreducible representations of the time-reversal group $\mathcal T$.}

 For magnetic systems there is one further restriction that must be considered: that of net magnetic moment. Within a crystallographic orbit, the system will prefer either a spin maximising or spin minimising state. The crystal may then be ferromagnetic (all orbits are spin maximising), antiferromagnetic (all orbits are spin minimising \textit{or} all orbits are spin maximising, but of equal and opposite magnitude), or ferrimagnetic (all orbits are spin maximising, with non-equal and opposite magnitude). For maximal epikernels with only one symmetry function, there is no degree of freedom here, however for maximal epikernels with \textit{multiple} symmetry functions, those symmetry functions returned by the projection formula may not be fully spin maximising or spin minimising as they are returned by the formula. Take, for instance, the two symmetry functions, $f_1$ and $f_2$, associated with the orbit of Mn atoms in the C$_{3i}$ maximal epikernel of IrMn$_3$:
\begin{align*}
    f_1 &= \text{Mn}_1:(1,1,0),\quad\text{Mn}_2:(1,0,1),\quad\text{Mn}_3:(0,1,1)\\
    f_2 &= \text{Mn}_1:(0,0,1),\quad\text{Mn}_2:(0,1,0),\quad\text{Mn}_3:(1,0,0)
\end{align*}
If we consider the total (normalised) spin moment of the two symmetry functions, $f_1$ has a net moment of $\frac{2}{3}$, while $f_2$ has a net moment of $\frac{1}{3}$. It can then be seen that $f_1+f_2$ (panel \textit{a)}, figure \ref{fig:IrMn_spins}) is a spin maximising symmetry function with a net spin of 1, while $f_1-2f_2$ (panel \textit{d)}, figure \ref{fig:IrMn_spins}) is a spin minimising symmetry function with a net spin of 0.

This consideration of net spin moment can be rigorously defined using the time reversal group of Bethe. $\mathcal T$, (table \ref{tab:double}) as systems with no net magnetic moment transform as the totally symmetric $A^+$ irrep, and systems with a net magnetic moment transform as the antisymmetric $A^-$ irrep. The same process of projecting the symmetry functions onto a new basis can be applied, with both $A^+$ and $A^-$ symmetry functions being targeted, thereby producing the full set of potential magnetic states (including non-magnetic)
\begin{table}[h!]
\centering
\begin{tabular}{|c|c|c|}
    \hline
    $\mathcal T$ & $E$ & $\bar{E}$ \\\hline
    $A^+$ & 1 & 1\\\hline
    $A^-$ & 1 & -1\\\hline
\end{tabular}
\caption{The magnetic doubling group $\mathcal T$ is applied to any standard space group to give rise to the corresponding double group(s). The operation $\bar{E}$ acts as identity on a non-spinor wavefunction, but applies a phase to a spinor wavefunction.}\label{tab:double}
\end{table}

\begin{acknowledgments}
The authors would like to thank Prof. Michael Bates of the University of York mathematics department for his generous conversations on the finer details of group theory. The Viking cluster was used during this project, which is a high performance compute facility provided by the University of York. We are grateful for computational support from the University of York, IT Services and the Research IT team.
R.A.L. acknowledges the computational resources provided by CSC (IT Center for Science, Finland) and financial support from the Research Council of Finland through the Quantum Flagship/QDOC doctoral pilot program.
\end{acknowledgments}

\newpage
\bibliography{bib}

\end{document}